\documentclass[sigconf]{acmart}
\AtBeginDocument{%
  }

\setcopyright{acmlicensed}
\copyrightyear{2018}
\acmYear{2018}
\acmDOI{XXXXXXX.XXXXXXX}

\acmConference[Conference acronym 'XX]{Make sure to enter the correct
  conference title from your rights confirmation email}{June 03--05,
  2018}{Woodstock, NY}
\acmISBN{978-1-4503-XXXX-X/2018/06}

\usepackage{amsmath}
\usepackage{mathtools}
\usepackage{amsthm}
\usepackage{amsfonts}
\usepackage{nicefrac}
\usepackage{microtype}
\usepackage{graphicx}
\usepackage{booktabs}
\usepackage{multirow}
\usepackage[table]{xcolor}
\usepackage[most]{tcolorbox}
\usepackage{xspace}
\usepackage{bm}
\usepackage{enumitem}
\usepackage{pifont}
\usepackage{longtable}
\usepackage{adjustbox}

\newcommand{\eg}{\textit{e.g.},\xspace}
\newcommand{\ie}{\textit{i.e.},\xspace}

\begin{document}
\settopmatter{printacmref=false} 
\title{Strategy-Aware Parameter-Efficient Adaptation for \\ LLM-based Auto-Bidding}

\author{Songyue Cai}
\authornote{Work is done
during the internship at Taobao \& Tmall Group
of Alibaba}
\email{sonnycai4649@gmail.com}
\affiliation{
  \institution{University of Electronic Science and Technology of China}
  \city{Chengdu}
  \country{China}
}
\affiliation{
  \institution{Taobao \& Tmall Group of Alibaba}
  \city{Beijing}
  \country{China}
}

\author{Lianyu Wang}
\email{wanglianyu.wly@taobao.com}
\affiliation{
  \institution{Taobao \& Tmall Group of Alibaba}
  \city{Beijing}
  \country{China}
}

\author{Shan Gu}
\email{gushan.gs@taobao.com}
\affiliation{
  \institution{Taobao \& Tmall Group of Alibaba}
  \city{Beijing}
  \country{China}
}

\author{Ziru Xu}
\email{ziru.xzr@taobao.com}
\affiliation{
  \institution{Taobao \& Tmall Group of Alibaba}
  \city{Beijing}
  \country{China}
}

\author{Jian Xu}
\email{xiyu.xj@taobao.com}
\affiliation{
  \institution{Taobao \& Tmall Group of Alibaba}
  \city{Beijing}
  \country{China}
}

\author{Xiaofeng Zhu}
\authornote{Corresponding authors.}
\authornote{Xiaofeng Zhu is the master's advisor of Songyue Cai.}
\email{zhuxf@hainanu.edu.cn}
\affiliation{
  \institution{Hainan University}
  \city{Haikou}
  \country{China}
}

\author{Bo Zheng}
\authornotemark[2]
\email{bozheng@alibaba-inc.com}
\affiliation{
  \institution{Taobao \& Tmall Group of Alibaba}
  \city{Beijing}
  \country{China}
}

\renewcommand{\shortauthors}{Cai et al.}
\newtcolorbox{promptbox}[1]{
    breakable,
    enhanced,
    colback=white,
    colframe=blue!40,
    colbacktitle=blue!15,
    coltitle=black,
    title=#1,
    fonttitle=\bfseries,
    arc=2mm,
    boxrule=0.6pt,
    left=2mm,
    right=2mm,
    top=1mm,
    bottom=1mm,
}

\begin{abstract}
Advertising bidding has evolved from manual strategies to auto-bidding systems better adapted for large-scale, dynamic auction environments. While recent advances in Large Language Models (LLMs) offer strong reasoning for auto-bidding, existing methods suffer from shallow trajectory-text interactions and require costly fine-tuning, hindering the efficient use of pretrained knowledge under diverse constraints. To address these challenges, we propose \textbf{SAGE}, a novel \underline{\textbf{S}}trategy-aware \underline{\textbf{A}}uto-bidding framework \underline{\textbf{G}}uided by LLMs for \underline{\textbf{E}}fficient bidding. SAGE introduces a parameter-efficient multi-modal alignment framework for constrained auto-bidding with LLMs. Specifically, SAGE comprises three key components: (i) the \textbf{position augmentation module} adopts temporal-semantic positional embeddings to effectively capture the intrinsic dynamics and semantic structures; (ii) the \textbf{text alignment module} leverages gated cross-attention to align the embedding spaces of trajectory and text modalities, enabling effective multi-modal fusion while alleviating the computational overhead caused by long trajectories; (iii) the \textbf{constraint-gated LoRA module} employs constraints as routing signals, activating only a small subset of experts to adapt the behavior of a frozen LLM efficiently. Extensive experiments on large-scale auto-bidding benchmark demonstrate that SAGE consistently achieves superior performance while tuning less than \textbf{10\%} of the trainable parameters required by full fine-tuning. Ablation studies further validate the critical contribution of each component to the framework's overall performance. The source code will be released at \textcolor[rgb]{0.7, 0.0, 0.125}{\url{https://github.com/YuzunoKawori/SAGE}}.
\end{abstract}


\maketitle

\section{Introduction}
Advertising bidding is a key decision-making problem in online advertising systems, where the goal is to determine optimal bids for billions of daily impression opportunities in a dynamic auction environment to achieve diverse advertisers’ economic objectives. With the rapid digitalization of commerce in recent years, the sheer volume of traffic and intensity of competition~\cite{wine2009internet} have rendered traditional manual bidding by human experts impractical~\cite{evans2009online,wang2015real}. Auto-bidding systems have become an important solution to the limitations of traditional manual bidding in high-traffic and highly competitive environments. By leveraging trained models, such systems can generate precise bids for large-scale ad traffic in real time based on current states and environmental information. Specifically, auto-bidding treats advertisers’ bidding decisions as a constrained sequential decision-making problem, where the goal of auto-bidding is to dynamically adjust bidding parameters throughout the delivery period to maximize conversions while satisfying advertiser-specified constraints such as cost-per-action (CPA)~\cite{he2021unified,mou2022sustainable}. As a result, auto-bidding has become a crucial mechanism for advertisers to achieve their business objectives in competitive advertising markets~\cite{ren2017bidding}.

\begin{figure*}[t!] 
\centering 
\includegraphics[width=\textwidth]{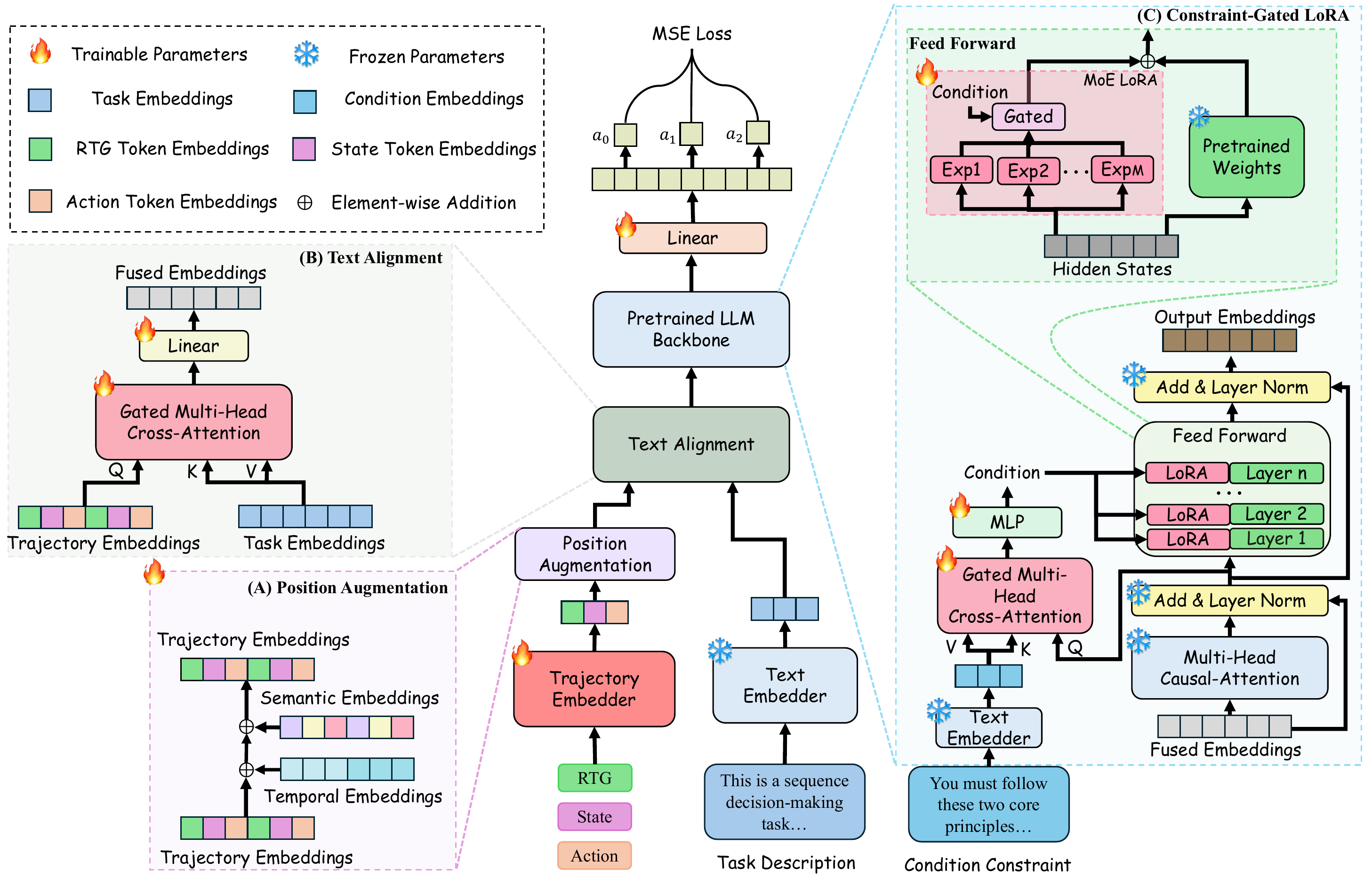} 
\vspace{-0.7em} 
\caption{The overall flowchart of the proposed SAGE. SAGE consists of three key modules (A, B, and C). Given an advertising bidding trajectory, SAGE first uses a trainable trajectory embedder to obtain the trajectory embeddings. The module (A) adds temporal–semantic positional embeddings to the trajectory embeddings. The module (B) then fuses the trajectory embeddings with the text embeddings obtained by encoding the task description through the frozen embedding layer of the LLM, to generate the fused embeddings. The module (C) first feeds the fused embeddings into the frozen pretrained LLM backbone to obtain intermediate hidden states from the causal attention layers. It then integrates the hidden states with the condition embeddings encoded from the condition constraint using the same frozen LLM embedding layer, to generate the condition gating signal. Based on this signal, Mixture-of-Experts (MoE) LoRA performs parameter-efficient adaptation on the frozen LLM backbone. Finally, a linear layer projects the last hidden states to produce the predicted action at the current timestep.}   
\label{fig:flowchart}
\vspace{-0.7em}
\end{figure*}

Existing auto-bidding methods can generally be divided into four categories, \ie offline reinforcement learning (RL) methods~\cite{kumar2020conservative,kostrikovoffline,mou2022sustainable,wen2022cooperative}, sequence modeling-based methods~\cite{chen2021decision, gao2025generative,lei2026generative,zhang2026generative}, planning-based generative bidding methods~\cite{gao2025segb, guo2024generative, janner2022planning, li2025generative}, and large language model (LLM)-based multi-modal methods~\cite{lv2026decisionllm,li2026lbm,zhu2026role}. Specifically, offline RL methods learn bidding policies from offline datasets by estimating state–action value functions and optimizing policies via Bellman backups. For example, CQL~\cite{kumar2020conservative} performs conservative value estimation for out-of-distribution actions to mitigate Q-value overestimation in offline RL, thereby improving the stability of policy learning. However, accurately estimating value functions in long-horizon and sparse-reward bidding environments remains challenging, often leading to unstable policy learning. To mitigate these challenges, researchers have turned to sequence modeling paradigms. Sequence modeling-based generative methods reformulate auto-bidding as a sequence modeling task and directly generate actions by modeling historical trajectories without explicitly learning value functions~\cite{ajayconditional}. For instance, decision transformer (DT)~\cite{chen2021decision} models return-to-go (RTG), states, and actions using a transformer~\cite{vaswani2017attention} architecture and autoregressively generates bidding actions conditioned on the desired return. Nevertheless, beyond trajectory generation, researchers have also explored explicit modeling of future dynamics and long-term planning in auto-bidding. Planning-based generative bidding methods further introduce planning mechanisms over the environment or future states on top of generative policies, guiding current bidding decisions by predicting future information. For example, DiffBid~\cite{guo2024generative} learns temporal periodic distributions to perform strategy planning and then generates actions through a control module. Since offline RL methods and generative methods primarily operate on structured trajectory data, emerging LLM-based multi-modal methods further enable advertisers’ constraints (\eg CPA) to be expressed through natural language. For example, LBM~\cite{li2026lbm} adopts a hierarchical framework where an LLM first generates a planning chain-of-thought under given bidding objectives and then produces actions guided by plan.

Despite the recent progress of LLM-based multi-modal methods, existing methods still face a fundamental challenge in achieving effective multi-modal alignment and strategy-aware parameter-efficient adaptation under complex constraints. This challenge mainly manifests in three aspects. First, in the trajectory representation stage, existing methods typically employ simple learnable positional embeddings to model sequential trajectories. Such designs mainly capture temporal dependencies, but overlook the semantic differences among token types, thus weakening the model’s ability to encode the structural information of bidding sequences. For example, DecisionLLM~\cite{lv2026decisionllm} assigns identical positional embeddings to different token types within the same timestep, making it difficult for the model to distinguish their semantic roles effectively. Second, in the multi-modal fusion stage, many methods rely on naive concatenation to fuse trajectory and text modalities before feeding them into the LLM, ignoring the embedding space discrepancy across modalities. This coarse-grained fusion strategy not only hinders effective cross-modal alignment but also incurs additional computational overhead. For instance, LBM~\cite{li2026lbm} directly concatenates trajectory, task description, and constraint for joint encoding, making it difficult to establish effective cross-modal interactions while also introducing excessively long input sequences. Finally, in the model adaptation stage, most existing methods rely on full fine-tuning to adapt LLMs to the auto-bidding task, requiring updates to a large number of parameters. This not only results in substantial training and deployment costs but also limits the model’s transferability to new tasks. For example, both LBM~\cite{li2026lbm} and DecisionLLM~\cite{lv2026decisionllm} adopt full fine-tuning strategies that update all parameters of the pretrained LLMs during training, resulting in substantial retraining costs when adapting to new tasks.

To address the above challenges, we propose a novel \underline{\textbf{S}}trategy-aware \underline{\textbf{A}}uto-bidding framework \underline{\textbf{G}}uided by LLMs for \underline{\textbf{E}}fficient bidding, referred to as \textbf{SAGE}. The overall framework is illustrated in Figure \ref{fig:flowchart} and consists of three key modules designed to systematically bridge the gaps in representation, fusion, and adaptation. Firstly, the \textbf{position augmentation module} disentangles positional embeddings into temporal and semantic components and assigns different positional embeddings to tokens across different timestep and semantic types, thereby enhancing the model’s understanding of trajectory structures while reducing additional parameters and addressing the first challenge. Secondly, the \textbf{text alignment module} performs efficient fusion between trajectory modality and text task description through cross-attention before encoding while aligning their embedding spaces, thereby improving the model’s understanding of text instructions and significantly shortening the input length and addressing the second challenge. Finally, the \textbf{constraint-gated LoRA module} freezes the pretrained backbone and updates only a small subset of parameters for efficient fine-tuning, while utilizing task constraints as gating signals to select different experts, thereby enabling adaptive strategy selection with low computational overhead and addressing the third challenge. Compared with previous methods, the main contributions of our work are summarized as follows:
\begin{itemize}
    \item We propose \textbf{SAGE}, a novel framework that makes LLM-based auto-bidding both strategy-aware and parameter-efficient, significantly improving decision-making with reduced computational overhead.
    \item We design the position augmentation module that enhances trajectory modeling by disentangling temporal and semantic positional information, enabling a deeper structural understanding with high parameter efficiency.     
    \item We introduce the text alignment module that effectively aligns trajectory and text embeddings while significantly compressing the input for the LLM.
    \item We design the constraint-gated LoRA module that enables parameter-efficient fine-tuning for auto-bidding by utilizing condition constraint as gating signals to dynamically adapt bidding strategies with minimal trainable parameters.
    \item We demonstrate that SAGE achieves superior performance on a large-scale advertising bidding benchmark while using 91.70\% fewer trainable parameters than full fine-tuning.
    
\end{itemize}

\section{Preliminaries}
\subsection{Problem Statement}
\label{statement}
In online advertising systems, thousands of advertisers compete for limited inventory by submitting bids over a time horizon comprising a chronological trajectory of $D$ ad opportunities. For each opportunity $i\in \{1,\ldots,D\}$, an advertiser submits a $bid_i$. During the auction process, the advertiser with the highest bid wins the impression, obtains the corresponding value $v_i$, and pays the associated cost $c_i$, which is typically determined by the second-price auction mechanism. The advertiser’s objective is to maximize the total value obtained from advertising impressions under a predefined budget constraint.
In many practical scenarios, advertisers may also impose additional constraints, such as CPA~\cite{he2021unified}. Therefore, an effective auto-bidding strategy should maximize the advertiser’s overall utility while satisfying these constraints. Formally, the objective of the auto-bidding problem can be expressed as:
\begin{equation}
\begin{aligned}
\max \quad & \sum_{i} o_i v_i \\
\text{s.t.} \quad & \sum_{i} o_i c_i \le B \\
& \frac{\sum_{i} c_{ij} o_i}{\sum_{i} p_{ij} o_i} \le C_j, \quad \forall j \\
& o_i \in \{0,1\}, \quad \forall i
\end{aligned}
\label{eq1}
\end{equation}
where $o_i$ denotes whether the advertiser wins the $i$-th impression, $v_i$ denotes the corresponding value, $B$ denotes the advertiser’s budget, $c_{ij}$ denotes cost associated with constraint $j$ for impression $i$, $C_j$ is the upper bound of $j$-th constraint provided by the advertiser, and $p_{ij}$ denotes the performance indicator, \eg conversions.

Previous studies have derived the theoretically optimal bidding formula under such constraints~\cite{he2021unified}:
\begin{equation}
bid_i^* = \lambda_0 v_i + \sum_{j=1}^{J} \lambda_j p_{ij} C_j,
\label{eq2}
\end{equation}
where $bid_i^*$ denotes the optimal bid for the $i$-th impression, $\lambda_0$ denotes the optimal bidding parameter, $J$ denotes the number of constraints, and $\lambda_j$ denotes the Lagrange multiplier for the $j$-th constraint. The objective is to dynamically adjust these parameters to optimize bidding decisions under multiple constraints.
\subsection{Sequential Decision for Auto-Bidding}
In dynamic auction marketplaces where both competitor strategies and opportunity distributions evolve continuously, static bidding parameters inevitably become suboptimal. This necessitates adaptive parameter adjustments based on real-time market feedback, fundamentally casting auto-bidding as a sequential decision-making problem. We formalize this process by defining the following interconnected components, where the model observes state $s_t$ to select action $a_t$, receives immediate reward $r_t$ as environmental feedback, and optimizes toward the cumulative RTG $R_t$: 
\begin{itemize}
    \item State $s_t$: a feature vector describing the bidding environment at timestep $t$, such as remaining budget, remaining time, and historical bidding information.
    \item Action $a_t$: the action taken at timestep $t$, represented as
    $a_t=(a^{\lambda_0}_t,a^{\lambda_1}_t,\ldots,a^{\lambda_J}_t)$.
    \item Reward $r_t$: the total value obtained from impressions won during timestep $t$. Given $D_t$ candidate impressions in this period, the reward is defined as $r_t = \sum_{i=1}^{D_t} o_i v_i$, where $D_t$ denotes the number of ad opportunities at the timestep $t$.
    \item RTG $R_t$: the cumulative reward from timestep $t$ until the terminal timestep $T$, defined as $R_t = \sum_{n=t}^{T} r_{n}$.
\end{itemize}
With these definitions, the entire bidding procedure can be represented as a trajectory $\tau = (R_0, s_0, a_0, \ldots, R_{T-1}, s_{T-1}, a_{T-1})$.

\subsection{LLM for Auto-Bidding}
LLM-based multi-modal methods have emerged as a promising direction for auto-bidding~\cite{lv2026decisionllm,li2026lbm}, leveraging pretrained LLM and text prompts to enhance task understanding and enforce conditional constraints (\eg CPA). Given a bidding decision trajectory $\tau$, previous LLM-based multi-modal methods~\cite{lv2026decisionllm,li2026lbm} first employ a lightweight trainable embedder $\mathcal{S}(\cdot)$ to encode the trajectory modality and obtain its embedding $\bm{e}=\mathcal{S}(\tau)\in \mathbb{R}^{L\times d}$, where $L=3T$ denotes the trajectory length and $d$ denotes the hidden dimension of the LLM. The trajectory embeddings consist of a sequence of embeddings across timestep, where timestep $t$ comprises three token types: RTG $R_t$, state $s_t$, and action $a_t$, formulated as:
\begin{equation}
    \bm{e_t}=\mathcal{S}(R_t,s_t,a_t), \quad \bm{e}=(\bm{e_0},\bm{e_1},\ldots, \bm{e_{T-1}}), \quad \bm{e_t}\in\mathbb{R}^{3\times d}.
    \label{eq3}
\end{equation}
Temporal information is injected by adding learnable positional embeddings $\bm{pos}=(\bm{pos_0},\ldots,\bm{pos_{T-1}})\in \mathbb{R}^{T\times d^\prime}$ uniformly to all token types $k \in \{R,s,a\}$ within each timestep:
\begin{equation}
\bm{e_{t,k}} = \text{Linear}(\text{Concat}(\bm{e_{t,k}}, \bm{pos_t})),
\label{eq5}
\end{equation}
where $\bm{e_{t,k}}\in\mathbb{R}^d$ denotes the embedding of the token of type $k$ at timestep $t$, $d^\prime$ denotes the dimension of positional embeddings, $\text{Concat}(\cdot)$denotes the concatenation operation. For the text modality, a text prompt $p$ is encoded via the frozen pretrained LLM embedder $\mathcal{T}(\cdot)$ into text embeddings $\bm{e_\text{text}}=\mathcal{T}(p)\in \mathbb{R}^{N\times d}$, where $N$ denotes the tokenized prompt length. Then, the text and trajectory embeddings are concatenated along the trajectory dimension to form the fused embeddings $\bm{x}$:
\begin{equation}
    \bm{x}=\text{Concat}(\bm{e_\text{text}}, \bm{e}), \bm{x}\in \mathbb{R}^{(N+L)\times d}.
    \label{eq6}
\end{equation}
Next, $\bm{x}$ is encoded into the hidden state $\bm{H}=\text{Backbone}(\bm{x})$ by the LLM backbone, where $\bm{H}\in\mathbb{R}^{(N+L)\times d}$, from which the state token $\bm{h^s_t}\in\mathbb{R}^d$ is linearly projected to predict the action:
\begin{equation}
\hat{a}_t = \text{Linear}(\bm{h_t^{s}}).
\label{eq7}
\end{equation}
Finally, the model is optimized via gradient descent using the mean squared error loss (MSE Loss):
\begin{equation}
\mathcal{L} = \frac{1}{T} \sum_{t=0}^{T-1} \| a_t - \hat{a}_t \|^2.
\label{eq8}
\end{equation}
\section{Methodology}
\subsection{Motivation}
Based on the above formulation, we further analyze the inherent limitations of existing LLM-based auto-bidding methods from the perspectives of trajectory representation, multi-modal fusion, and model adaptation. Specifically, in trajectory representation, existing methods inject temporal information by uniformly adding learnable positional embeddings to all token types within the same timestep according to Eq. (\ref{eq5}). Such a design implicitly assumes semantic homogeneity across RTG, state, and action tokens, making it difficult for the model to distinguish their distinct functional roles and thereby limiting its ability to capture the hierarchical structure of bidding trajectories. Moreover, the reliance on learnable embeddings introduces additional parameters and necessitates retraining when trajectory lengths vary, constraining transferability across different bidding horizons.

In multi-modal fusion, previous methods directly concatenate trajectory embeddings with text embeddings according to Eq. (\ref{eq6}). This coarse-grained fusion strategy treats trajectory data and textual instructions as semantically aligned by default, overlooking the inherent embedding space discrepancy and the semantic hierarchy between task description and condition constraint. As a result, it fails to establish effective cross-modal interactions while also introducing additional text embeddings for the LLM backbone to encode, leading to inefficient multi-modal fusion.

In model adaptation, previous methods optimize the model by fully fine-tuning the LLM backbone according to Eq. (\ref{eq8}). As a result, the rich pretrained prior knowledge may be diluted. This paradigm also incurs substantial computational costs and limits the model’s adaptability when transferred to new advertising scenarios. These intertwined issues collectively motivate our core research question: \textit{How can we achieve strategy-aware, parameter-efficient adaptation of LLMs for auto-bidding while preserving structural fidelity, cross-modal alignment, and pretrained knowledge?}

To address these limitations, we propose a novel LLM-based auto-bidding framework with three components: position augmentation, text alignment, and constraint-gated LoRA modules. The position augmentation module (Section \ref{pos}) tackles the entangled trajectory embeddings, the text alignment module (Section \ref{alignment}) resolves the inefficient cross-modal fusion, and the constraint-gated LoRA module (Section \ref{lora}) overcomes the costly adaptation paradigm.

\subsection{Position Augmentation}
\label{pos}
Previous methods typically design positional embeddings for trajectory tokens by only considering temporal differences across timestep, while overlooking that auto-bidding trajectories naturally contain multiple tokens with different semantics within the same timestep. Moreover, these methods usually adopt learnable positional embeddings by assigning a trainable parameter to each timestep. When the input trajectory length changes, such encodings often become ineffective.\\
\textbf{Semantic embeddings.}~~To address this issue, we design a position augmentation module. Specifically, a typical auto-bidding trajectory contains three tokens at each timestep: RTG $R_t$, state $s_t$, and action $a_t$. These tokens represent different semantic information and provide distinct perspectives for the model during decision-making. In addition, existing LLM-based methods often explicitly emphasize the names and semantics of different tokens when constructing text prompts~\cite{li2026lbm}. Therefore, ignoring semantic differences when designing positional embeddings may lead to incomplete information representation and further weaken the LLM’s ability to understand auto-bidding trajectories. Based on this observation, it is necessary to design distinct positional embeddings for tokens with different semantics. To achieve this, we introduce only a small number of trainable parameters $\bm{pos^s}=(\bm{pos^s_{R}},\bm{pos^s_s}, \bm{pos^s_a}), \bm{pos^s} \in \mathbb{R}^{3\times d}$ to represent semantic embeddings for each token type, thereby enhancing the semantic information within the trajectory. Formally, this process can be described by the following equation:
\begin{equation}
\bm{e_{t,k}} = \bm{e_{t,k}} + \bm{pos^s_k}, \quad k \in \{R,s,a\}.
\label{eq9}
\end{equation}
Through Eq. (\ref{eq9}), we assign consistent semantic embeddings to tokens with the same semantic type across different timestep, which enhances the distinction between different semantic tokens and improves the LLM’s understanding of trajectory information. \\
\textbf{Temporal embeddings.}~~Furthermore, we consider the design of temporal embeddings. Previous methods typically use a trainable parameter $\bm{pos}\in \mathbb{R}^{T\times d}$ as the temporal embeddings. However, in real-world environments, bidding trajectories often have varying numbers of timestep. In such cases, this method not only introduces additional trainable parameters but also lacks effective scalability when the trajectory length changes. To address this issue, we adopt a non-parametric positional encoding scheme, namely sinusoidal positional encoding~\cite{vaswani2017attention}, to obtain temporal embeddings $\bm{pos^\prime}=(\bm{pos^\prime_0},\bm{pos^\prime_1},\ldots, \bm{pos^\prime_{T-1}}), \bm{pos^\prime} \in \mathbb{R}^{T\times d}$. Finally, we replace the original positional embeddings with the proposed position augmentation module. Formally, it can be written as follows:
\begin{equation}
    \bm{e_{t,k}} = \bm{e_{t,k}} +\bm{pos^\prime_t}+ \bm{pos^s_k}.
    \label{eq10}
\end{equation}
By replacing Eq. (\ref{eq5}) with Eq. (\ref{eq10}), we model the missing semantic positional information in previous methods with few trainable parameters, while adopting non-parametric temporal embeddings to improve scalability. By jointly modeling semantic and temporal information, the module enhances the distinction among tokens with different semantics and supports variable-length trajectories. As a result, it improves the LLM’s understanding of trajectory information and addresses the first limitation of previous methods.
\subsection{Text Alignment}
\label{alignment}
In Section \ref{pos}, we introduce a position augmentation module that enhances trajectory embeddings by incorporating semantic embeddings and improved temporal embeddings, enabling LLMs to better understand trajectory modality and facilitating multi-modal fusion with text information. However, existing methods typically rely on simple concatenation for multi-modal fusion, which fails to adequately capture complex interactions between trajectory and text modalities and makes it difficult to effectively align their heterogeneous feature spaces without full fine-tuning, potentially disrupting pretrained prior knowledge. At the same time, auto-bidding prompts often contain a detailed task description and condition constraint, substantially increasing the input trajectory length and introducing considerable computational overhead.


Motivated by these observations, we propose the text alignment module to achieve efficient and effective multi-modal fusion. Inspired by the success of cross-attention mechanisms~\cite{vaswani2017attention} in information retrieval and alignment tasks~\cite{li2023blip,jin2024time,rombach2022high,alayrac2022flamingo,gheini2021cross,liu2024towards}, we introduce a gated cross-attention where trajectory embeddings serve as the query and text prompt embeddings act as the key and value. In cross-attention, the query representation retrieves relevant semantic information from the key-value pairs while preserving the original sequence structure of the query modality, enabling flexible cross-modal interaction. Hence, using trajectory embeddings as queries allows the model to enhance trajectory representations with task semantics while maintaining compatibility with existing trajectory prediction paradigms. In addition, it is worth noting that the complex condition constraint is difficult to interpret directly at the embedding level. Therefore, the text prompt used in this module mainly describes the task semantics to enhance trajectory embeddings, while deeper interactions with constraint conditions will be handled in Section \ref{lora}.

Accordingly, the proposed text alignment module aligns the trajectory and text modalities to obtain the fused embeddings. The formulation is given as follows:
\begin{equation}
    \bm{x}=\text{Linear}(\text{Gated\_attn}(Q=\bm{e},K=\bm{e}_{\text{task}},V=\bm{e}_{\text{task}})), \bm{x}\in\mathbb{R}^{L\times d},
    \label{eq11}
\end{equation}
where $Q$, $K$, and $V$ represent the query, key, and value embeddings, $\text{Gated\_attn}(Q,K,V)$  denotes a gated cross-attention layer, in which the query embeddings attend to the key and value embeddings to produce cross-modal interaction features, and a learnable gating function with sigmoid activation is applied to control how much of these features are preserved in the fused embeddings, $\bm{e_\text{task}}=\mathcal{T}(p_\text{task})$, where $\bm{e_\text{task}}\in \mathbb{R}^{N_\text{task}\times d}$ denotes the task embeddings, $N_\text{task}$ denotes the length of the tokenized task description prompt $p_\text{task}$. This gating mechanism is introduced to enable more flexible and effective multi-modal fusion.
 
By replacing Eq. (\ref{eq6}) with Eq. (\ref{eq11}), we leverage the cross-attention mechanism to enable trajectory embeddings to adaptively extract relevant information from text embeddings, thereby naturally aligning the two modalities in the embedding space. Notably, this fusion process preserves the original trajectory length without introducing additional text tokens. As a result, the proposed module achieves effective multi-modal fusion and embedding alignment while avoiding the computational overhead caused by lengthy inputs, effectively addressing the second limitation of previous methods.

\subsection{Constraint-Gated LoRA}

\label{lora}
Previous methods typically rely on full fine-tuning to adapt the LLM backbone for auto-bidding, since the simple fusion strategy cannot effectively bridge the embedding space discrepancy between trajectory and text modalities. This paradigm incurs substantial computational overhead and limits the transferability of the model to new bidding scenarios and constraint settings. In Section \ref{alignment}, the proposed text alignment module achieves effective cross-modal alignment through semantic interaction, thereby enabling parameter-efficient fine-tuning for model adaptation. 

To do this, we propose the constraint-gated LoRA module, which performs parameter-efficient fine-tuning on the frozen LLM backbone through a mixture-of-experts (MoE) LoRA. LoRA~\cite{hu2022lora} introduces trainable low-rank adapters into the original weight matrices and has demonstrated strong performance in various transfer learning tasks~\cite{yang2025mtl,nejatimanzari2026sparse,luo2024moelora,dettmers2023qlora}, achieving performance comparable to full fine-tuning while updating a small number of parameters. Therefore, LoRA provides an effective solution for efficient adaptation in auto-bidding scenarios. 

However, advertisers adopt different bidding strategies under diverse optimization objectives and constraints, such as maximizing conversions under CPA constraints, making it difficult for a single LoRA adapter to capture heterogeneous bidding behaviors. Inspired by MoE methods~\cite{luo2024moelora,zhang2025more}, we introduce multiple strategy-specific LoRA experts and employ condition-aware routing to dynamically select appropriate experts according to advertiser constraints. To do this, we first introduce a condition constraint prompt $p_\text{con}$ to describe advertiser objectives and constraints, and employ a gated cross-attention mechanism to enable the trajectory to extract relevant information from the condition constraint for expert selection. The formulation can be written as follows:
\begin{equation}
    \bm{C^l}=\text{Gated\_attn}(Q=\bm{H^l},K=\bm{e_\text{con}},V=\bm{e_\text{con}}), \ \ \ \ \bm{C^l} \in \mathbb{R}^{L\times d},
\end{equation}
where
\begin{equation}
     \bm{C^l}=(\bm{c^l_{0,k=R}}, \bm{c^l_{0,k=s}}, \bm{c^l_{0,k=a}} \ldots,\bm{c^l_{T-1,k=a}}),
\end{equation}
where $\bm{H}^l$ denotes the hidden states after the causal self-attention, residual connection, and layer normalization at the $l$-th transformer layer, $\bm{C^l}$ denotes the constraint information extracted by the trajectory at the $l$-th layer, and $\bm{e}_{\text{con}} = \mathcal{T}(p_{\text{con}})$, where $\bm{e}_{\text{con}} \in \mathbb{R}^{N_\text{con}\times d}$ denotes the condition embeddings, $N_\text{con}$ is the length of the tokenized condition constraint prompt. Then, we employ a multi-layer perceptron $\text{MLP}(\cdot)$ to aggregate the information extracted by different semantic tokens within the same timestep, which serves as the input to the MoE gating mechanism. The formulation is as follows:
\begin{equation}
    \bm{y^l_{t}}=\text{MLP}([\bm{c^l_{t,k=R}}, \bm{c^l_{t,k=s}}, \bm{c^l_{t,k=a}}]), \ \ \ \ \bm{y^l_{t}} \in \mathbb{R}^d,
\end{equation}
where $\bm{y^l_{t}}$ denotes the condition gating signal at timestep $t$ in the $l$-th layer. Since $\bm{y^l_{t}}$ encodes the constraint-aware information of the current token, we use it as the input to the MoE-LoRA gated network to dynamically select the appropriate LoRA expert for each trajectory token, thereby enabling adaptive strategy selection. The formulation is as follows:
\begin{equation}
\left\{
\begin{aligned}
\bm{g^l_{t}} &= \text{Softmax}(\text{Router}^l(\bm{y^l_{t}})) \\
m^l_t &= \arg\max_m \bm{g_{t}^{l}},
\end{aligned}
\right.
\end{equation}
where $\text{Router}^l(\cdot)$ denotes the gating network at $l$-th layer, $\bm{g^l_{t}}\in\mathbb{R}^{M}$ denotes the routing weight over experts at $l$-th layer and timestep $t$, $M$ denotes the number of experts, and $m^l_t$ denotes the index of the selected expert at $l$-th layer and timestep $t$. After selecting the strategy expert, we adopt the standard LoRA scheme~\cite{hu2022lora} to perform low-rank adaptation on the backbone network. Specifically, the output of the linear layer is computed as the sum of the original backbone output and the increment produced by the low-rank adapter. The formulation is given as follows:
\begin{equation}
\bm{\tilde{H}_{t}^{l}} = \bm{H_{t}^{l}}\bm{W^\top}  + \frac{\alpha}{r}\bm{H_{t}^{l}}\bm{A^\top_{m^l_t}}\bm{B^\top_{m^l_t}} ,
\label{eq16}
\end{equation}
where $\bm{\tilde{H}}_{t}^{l}\in\mathbb{R}^{3\times d}$ denotes the updated hidden state after the transformation, $\bm{H}_t^l \in \mathbb{R}^{3 \times d}$ denotes the hidden state at timestep $t$ of $\bm{H}^l$ in the $l$-th layer, $\bm{W}$ denotes the base weight matrix of the linear transformation, $\alpha$ denotes the LoRA scaling factor, $r$ denotes the rank of the low-rank adaptation, where $r \ll d$ to ensure parameter-efficient adaptation, $\bm{A_{m^l_t}}$ denotes the low-rank down-projection matrix, and $\bm{B_{m^l_t}}$ denotes the low-rank up-projection matrix. 

Through Eq. (\ref{eq16}), domain adaptation can be achieved by training only a small number of parameters while keeping the backbone frozen. Furthermore, the constraint-gated LoRA module models the interaction between trajectory information and constraint condition to generate timestep-wise gating signals, and leverages a MoE LoRA to enable efficient fine-tuning and flexible policy selection, thereby addressing the third limitation of previous methods. 

Overall, the three proposed modules jointly address the three aforementioned key challenges.
\section{Experiments}

\subsection{Experimental Details}
\begin{table*}[ht]
\caption{Score comparison between our proposed method and previous baselines on the AuctionNet benchmark under different budget scales. Bold indicates the best performance, and underlined values denote the second-best performance.}
\vspace{-0.7em} 
\label{tab:1}
\centering
\setlength{\tabcolsep}{6pt}
\resizebox{0.85\textwidth}{!}{
\begin{tabular}{l|c|ccccccccc|c}
\toprule
\textbf{Dataset} & \textbf{Budget} & \textbf{USCB} & \textbf{BCQ} & \textbf{CQL} & \textbf{IQL} & \textbf{DiffBid} & \textbf{DT} & \textbf{CDT} &\textbf{DT-score}& \textbf{SAGE} & \textbf{Improve} \\
\midrule

\multirow{5}{*}{AuctionNet-dense}
& 50\%  & 86 & 190 & 113 & 164 & 54 &\underline{191}  &174 &178 & \textbf{204}&6.81\%  \\
& 75\%  & 135 & 259 &139 & 232 & 100 &265  &242 &\underline{268} &\textbf{274} &2.24\%  \\
& 100\% & 157 & 321 &171 & 281 &152  &329  &326 &\underline{334}& \textbf{349}& 4.49\% \\
& 125\% & 220 & 379 &201 & 355 &193  &\underline{396}  &378 &395 &\textbf{403} & 1.77\% \\
& 150\% & 281 & 429 &238  &401 &234  &\underline{450}  &433 &441 &\textbf{462}& 2.67\% \\

\midrule

\multirow{5}{*}{AuctionNet-sparse}
& 50\%  &11.5  &17.7  &12.8  &16.5  &9.9  &14.8  &11.2 &\underline{17.7} &\textbf{19.6}& 10.73\%\\
& 75\%  &14.9  &24.6  &16.7  &22.1  &15.4  &22.9  &18.0 &\underline{25.9}  &\textbf{29.7} &14.67\% \\
& 100\% &17.5  &31.1  &22.2  &30.0  &19.5  &29.6  &31.2 &\underline{33.2}  &\textbf{35.8} & 7.83\%\\
& 125\% &26.7  &34.2  &28.6  &37.1  &25.3  &34.3  &31.7 &\underline{39.6}  &\textbf{40.8} &3.03\%\\
& 150\% &31.3  &37.9  &35.8  &43.1  &30.8  &44.5  &39.1 &\underline{44.7} & \textbf{46.5}&4.03\%\\

\bottomrule
\end{tabular}
}
\vspace{-0.7em}
\end{table*}

\begin{table*}[ht]
\caption{Score and conversions comparison between our proposed method and LLM-based multi-modal methods on the AuctionNet benchmark under different budget scales. Bold indicates the best performance, and underlined values denote the second-best performance.}
\vspace{-0.7em} 
\label{tab:2}
\centering
\resizebox{0.85\textwidth}{!}{
\begin{tabular}{l|l|cccccc|c}
\toprule
\textbf{Dataset} & \textbf{Metric} & 
\textbf{Prompting} & \textbf{SFT} & \textbf{GRPO} & \textbf{LLM-DT} & \textbf{Prompt-LLM-DT} & \textbf{SAGE} & \textbf{Improve} \\
\midrule

\multirow{2}{*}{AuctionNet-dense}
& Conversions ($\uparrow$)
&289  & 297 &327  &\underline{366}  &355  &\textbf{371}  & 1.37\% \\

& Score ($\uparrow$)
&286  & 286 & 292 &\underline{328}  &322  & \textbf{349} & 6.40\% \\

\midrule

\multirow{2}{*}{AuctionNet-sparse}
& Conversions ($\uparrow$)
& 24.2 &27.0  &29.9  &\underline{36.0}  &35.8  &\textbf{38.3}  & 6.39\% \\

& Score ($\uparrow$)
&23.4  &26.9  & 29.7 &\underline{32.5}  & 30.3 &\textbf{35.8}  & 10.15\% \\

\bottomrule
\end{tabular}}
\vspace{-0.7em}
\end{table*}
\textbf{Datasets.}~~To evaluate the effectiveness of our method, we use AuctionNet~\cite{su2024a}, a large-scale advertising bidding dataset constructed by Alibaba. The dataset contains two variants, namely a dense version and a sparse version, referred to as AuctionNet-dense and AuctionNet-sparse. Each dataset includes data from 48 advertisers, with more than 500,000 impression opportunities in each period. More details about the dataset are provided in Appendix \ref{dataset}.\\
\textbf{Comparison methods.}~~The comparison methods include four offline RL methods (\ie USCB~\cite{he2021unified}, BCQ~\cite{fujimoto2019off}, CQL~\cite{kumar2020conservative}, IQL~\cite{kostrikovoffline}), one planning-based generative bidding method (\ie DiffBid~\cite{guo2024generative}), and three sequence modeling-based methods (DT~\cite{chen2021decision}, CDT~\cite{liu2023constrained}, DT-score), where DT-score further fine-tunes DT using Q-values~\cite{li2025gas}. We also compare with five LLM-based methods (\ie Prompting, SFT, GRPO~\cite{shao2024deepseekmath}, LLM-DT, Prompt-LLM-DT) for decision-making. Specifically, Prompting, SFT, and GRPO operate within the language modality, while LLM-DT and Prompt-LLM-DT process numerical sequences to generate numerical actions, with Prompt-LLM-DT further incorporating a language task description~\cite{li2026lbm}.\\
\textbf{Implementation details.}~Except for the ablation studies, our method is built upon the pretrained Qwen2.5-Instruct-0.5B model~\cite{bai2023qwen}. During training, the backbone parameters and the text embedder remain frozen. The batch size is 128, the learning rate is 1e-5, and AdamW~\cite{loshchilov2018decoupled} is used as the optimizer. The length of the trajectory is 10, and the maximum number of training steps is 400,000. More implementation details are provided in Appendix \ref{implementation}.\\
\textbf{Evaluation metrics.}~~The evaluation metrics mainly include the following two measures:
\begin{itemize}
    \item Conversions: the total number of conversions within one advertising period without constraints, defined as $\sum_i o_i v_i$.
    \item Score: the value under CPA constraints. A penalty defined as  $\min\left\{\left(\frac{1}{\text{ratio}}\right)^2,1\right\}$  
is applied, where $\text{ratio}$ is the ratio between the real CPA and the target CPA. The score is defined as $\sum_i o_i v_i \times \text{penalty}$.
\end{itemize}
\subsection{Main Results}

We report the experimental results of all methods on the AuctionNet benchmark, including its two variants, AuctionNet-dense and AuctionNet-sparse. Table \ref{tab:1} compares our proposed method with traditional auto-bidding methods, while Table \ref{tab:2} compares our method with several LLM-based multi-modal methods.\\
\textbf{Comparison to traditional baselines.}~~Our method achieves the best results, followed by DT-score, DT, CDT, BCQ, IQL, CQL, USCB, and DiffBid, in terms of different budget scales on all variants. For instance, under the 100\% budget setting, our method improves by 4.49\% and 7.83\%, respectively, compared to the best comparison method (\ie DT-score) on the AuctionNet-dense and AuctionNet-sparse variant, and by 129.60\% and 83.59\% compared to the worst comparison method (\ie DiffBid) in terms of score. This performance improvement may stem from the introduction of multi-modal information in our method, which enables the model to leverage the rich pretrained prior knowledge of LLMs. Moreover, by designing prompts to explicitly express and reinforce constraints, the model is guided to learn decision strategies that better satisfy these constraints, thereby achieving improved performance.\\
\textbf{Comparison to LLM-based methods.}~~Our method achieves the best results, followed by LLM-DT, Prompt-LLM-DT, GRPO, SFT, Prompting, in terms of different budget scales on all variants. For instance, under the 100\% budget setting, our method improves by 6.40\% and 10.15\%, respectively, compared to the best comparison method (\ie LLM-DT) on the AuctionNet-dense and AuctionNet-sparse variants, and by 22.03\% and 52.99\% compared to the worst comparison method (\ie Prompting) in terms of score. One possible reason is that our method introduces distinctive semantic embeddings for the bidding trajectory, which enhances the model’s understanding of different types of tokens. Notably, our method achieves larger improvements on the sparse variants, which may be attributed to the proposed text alignment module that leverages a cross-attention mechanism to enable more effective fusion and alignment between the trajectory modality and the text modality. As a result, our method achieves better performance compared with previous LLM-based methods. Moreover, it is also worth noting that LLM-based methods generally achieve strong performance, which further indicates that incorporating LLMs and utilizing sufficient  rich pretrained prior knowledge can effectively improve the bidding decision capability in auto-bidding tasks.
\begin{table}[h]
\caption{Ablation studies on position augmentation and text alignment.}
\vspace{-0.7em} 
\label{tab:3}
\centering
\resizebox{\columnwidth}{!}{
\begin{tabular}{lc}
\toprule
\textbf{Method} 
& \textbf{Score$\uparrow$} \\
\midrule
\textbf{w/o position augmentation} & 243  \\
\textbf{w/o text alignment} & 324  \\
\textbf{w/o position augmentation and text alignment} & 212  \\
\rowcolor{gray!20}
\textbf{SAGE} & \textbf{349}  \\
\bottomrule
\end{tabular}
}
\vspace{-0.7em}
\end{table}
\subsection{Ablation Studies}
\textbf{Effectiveness of position augmentation and text alignment.}~~The proposed method involves three key components, \ie the position augmentation, the text alignment, and the constraint-gated LoRA. Since the ablation of the constraint-gated LoRA module involves modifications to both the fine-tuning strategy and the text prompt, we defer the analysis of its effectiveness to a later section. In this section, we first evaluate the effectiveness of the position augmentation and text alignment modules. To this end, we report the score performance on the AuctionNet benchmark after ablating different components in Table \ref{tab:3}. The experimental results demonstrate that each component of our framework is effective. Specifically, compared with the full model, removing the position augmentation module leads to a decrease of 30.37\% in score, indicating that introducing distinctive semantic embeddings for trajectory information helps strengthen the distinction between different types of tokens and thus improves the model’s discriminative capability. When the text alignment module is removed, the score drops by 7.16\%, suggesting that leveraging cross-attention to enable interaction between the trajectory modality and the text modality can effectively facilitate cross-modal fusion and alignment, thereby improving the model’s understanding of the auto-bidding task. When both modules are removed, the performance further decreases by 39.26\%, which indicates that the proposed components work collaboratively and jointly contribute to achieving the best performance.\\
\noindent\textbf{Effectiveness of different positional embeddings.}~~Since the proposed position augmentation module consists of both semantic and temporal embeddings, we further analyze the contribution of each component and compare them with the conventional method that assigns learnable positional embeddings to each timestep. As shown in Table \ref{tab:4}, we report the score performance and parameter counts  on the AuctionNet benchmark when using temporal embeddings, semantic embeddings, and traditional positional embeddings, respectively. 
\begin{table}[h]
\vspace{-0.7em} 
\caption{Experiments on the impact of different positional embeddings, $\text{dim}$ denotes the hidden dimension, $\text{dim}^\prime$ denotes the dimension of traditional position embeddings, and $\text{timestep}$ denotes the total timestep in the trajectory.}
\vspace{-0.7em} 
\label{tab:4}
\centering
\setlength{\tabcolsep}{6pt}
\resizebox{\columnwidth}{!}{
\begin{tabular}{lcc}
\toprule
\textbf{Method} 
& \textbf{Score $\uparrow$}
& \textbf{Parameter counts $\downarrow$}\\
\midrule
\textbf{w/o semantic embeddings} & 249 & 0\\
\textbf{w/o temporal embeddings} & 315 & $3\times \text{dim}$ \\
\textbf{traditional position embeddings} & 250& $\text{timestep}\times \text{dim}^\prime+(\text{dim}+\text{dim}^\prime)\times \text{dim}$ \\
\rowcolor{gray!20}
\textbf{SAGE} & \textbf{349} &$3\times \text{dim}$ \\
\bottomrule
\end{tabular}
}
\vspace{-0.7em}
\end{table}
The results show that introducing semantic embedding significantly improves performance, demonstrating the effectiveness of the proposed design. Moreover, compared with traditional learnable positional embeddings, the sinusoidal temporal embeddings achieve comparable performance without introducing additional parameters and remains effective when the number of timestep changes, 
highlighting its efficiency and scalability.\\
\noindent\textbf{Effectiveness of constraint-gated LoRA.}~~To evaluate the effectiveness of the proposed constraint-gated LoRA module, we conduct ablation studies from two perspectives: prompt design and fine-tuning strategy. Since directly removing the module simultaneously affects both the utilization of constraint information and the parameter adaptation mechanism, we design more fine-grained comparisons. Specifically, after removing LoRA, we consider two training strategies: full fine-tuning and frozen parameters. Meanwhile, for prompt design, we compare removing the constraint information with concatenating it to the task description. The results are reported in Table \ref{tab:5}. The results show that preserving constraint information consistently improves performance, indicating that it provides important guidance for bidding decisions. In addition, freezing the backbone parameters performs better than full fine-tuning. A possible reason is that full fine-tuning a large language model may disrupt its rich pretrained prior knowledge, whereas keeping the backbone frozen helps retain its general capabilities and leads to more stable decision-making. Overall, the proposed constraint-gated LoRA module effectively leverages constraint information through a gating mechanism and performs efficient parameter adaptation via LoRA, which together contribute to improved performance. In addition to the superior performance, compared with full fine-tuning, our method uses less than 10\% of the trainable parameters while achieving effective LLM adaptation. These results demonstrate both the efficiency and effectiveness of the proposed method.\\
\begin{table}[h]
\caption{Ablation studies on the effectiveness of the constraint-gated LoRA module.}
\vspace{-0.7em} 
\label{tab:5}
\centering
\setlength{\tabcolsep}{6pt}
\resizebox{\columnwidth}{!}{
\begin{tabular}{cccccc}
\toprule
\textbf{LoRA} & \textbf{Condition} & \textbf{Full FT} 
& \textbf{Score$\uparrow$} & \textbf{Trainable Params (M)$\downarrow$} \\
\midrule

\ding{55} & \ding{55} & \ding{55} & 236 & 22.7 \\
\ding{55} & concat & \ding{55} & 334 & 22.7 \\
\ding{55} & \ding{55} & \ding{51} & 144 & 380.6 \\
\ding{55} & concat & \ding{51} & 173 & 380.6 \\
\ding{51} & gated & \ding{55} & \textbf{349} & 31.6 \\

\bottomrule
\end{tabular}
}
\vspace{-0.7em}
\end{table}
\begin{table}[h]
\caption{Experiments on AuctionNet benchmark with different backbone architectures.}
\vspace{-0.7em} 
\label{tab:6}
\centering
\setlength{\tabcolsep}{6pt}

\resizebox{\columnwidth}{!}{
\begin{tabular}{lcccc}
\toprule
\textbf{Method} 
& \multicolumn{2}{c}{\textbf{AuctionNet-dense}} 
& \multicolumn{2}{c}{\textbf{AuctionNet-sparse}} \\
\cmidrule(lr){2-3} \cmidrule(lr){4-5}
& \textbf{Score$\uparrow$} 
& \textbf{Conversions$\uparrow$}
& \textbf{Score$\uparrow$} 
& \textbf{Conversions$\uparrow$} \\
\midrule
\textbf{Qwen2.5-Instruct-0.5B} & \textbf{349} & \textbf{371} & \textbf{35.8} & \textbf{38.3} \\
\textbf{Qwen2.5-Instruct-1.5B} & 347 & 370 & 33.4 & 35.6  \\
\textbf{Qwen2.5-Instruct-3B} & 344 & 369 & 35.0 & 36.5 \\
\bottomrule
\end{tabular}
}
\vspace{-0.7em}
\end{table}
\begin{figure}[h] 
    \centering 
    \vspace{-0.7em} 
    \includegraphics[width=\columnwidth]{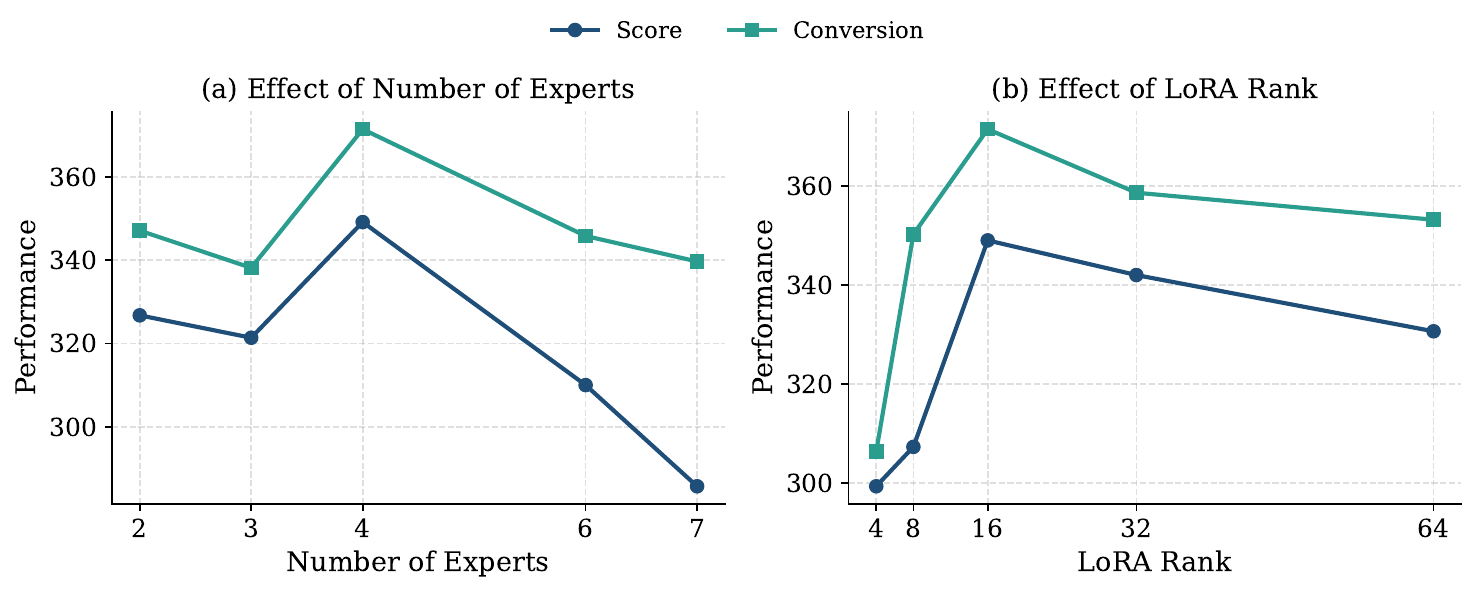} 
    \caption{Experiments on AuctionNet benchmark with different hyperparameter settings.}
    \vspace{-1.4em} 
    \label{fig:hyper}
\end{figure}
\noindent\textbf{Scalability across different backbone models.}~~To evaluate the scalability of our method, we conduct experiments on different backbone models. The results reported in Table \ref{tab:6} indicate that our method consistently achieves strong performance across all architectures, demonstrating its effectiveness and scalability. We also observe a slight performance drop on larger models. One possible reason is that adapting larger models becomes more challenging under limited training data, making it harder for the parameters to converge effectively. In addition, prior studies suggest that more precise representations do not necessarily lead to better task performance~\cite{zhu2026role}; instead, the alignment between model representations and the target task also plays an important role.\\
\noindent \textbf{Hyperparameter sensitivity analysis.}~~To analyze the impact of the number of MoE experts $M$ and the LoRA rank $r$ on model performance, we report the results under different configurations in Figure \ref{fig:hyper}. The results show that our method achieves strong and stable performance across a certain range of settings by $M \in \{2,3,4\}$ and $r\in\{16,32,64\}$. Notably, when the number of experts becomes large (\eg $M=7$), the performance drops noticeably. This may be because increasing the number of experts makes it more difficult for the routing network to learn stable assignment patterns, which can lead to imbalanced or inaccurate expert selection and thus degrade the overall performance. In addition, performance also decreases when the LoRA rank is too small (\eg $r=4$), since a low rank restricts the expressive capacity of parameter updates, making it harder for the model to adapt to the task.\\
\noindent \textbf{MoE load balancing analysis.}~~We further analyze the MoE load balancing in Appendix \ref{moe}.
\section{Conclusion}
In this paper, we propose \textbf{SAGE}, a new LLM-guided framework for auto-bidding. Specifically, we first design a position augmentation module to enhance trajectory modeling with temporal–semantic positional embeddings. We then introduce a text alignment module to align trajectory and text information through cross-attention. Finally, we propose a constraint-gated LoRA module that uses constraints as gating signals to select experts while enabling parameter-efficient fine-tuning of pretrained LLMs. Experimental results demonstrate the effectiveness and efficiency of our method.
\section*{Acknowledgments}
This work was supported by Alibaba Group through Alibaba Research Intern Program.
\bibliographystyle{ACM-Reference-Format}
\bibliography{sample-base}

\clearpage

\appendix
\section{APPENDIX}
\subsection{Related Work}
\label{related work}
\textbf{Auto-bidding methods.}~~With the rapid growth of e-commerce platforms, auto-bidding for online advertising has become an important research topic. Early studies mainly relied on control-based methods~\cite{realtimechen,NIPS2017_da0d1111,yang2019bid,zhang2016feedback}, such as PID~\cite{realtimechen} and OnlineLP~\cite{NIPS2017_da0d1111}, which adjusted bidding strategies through manually designed optimization or control rules. However, as auction environments become increasingly dynamic and the scale of advertising traffic continues to grow, these methods face significant challenges in scalability and adaptability. Benefiting from the rapid development of RL and strong baseline models, recent auto-bidding methods can generally be categorized into four groups: traditional offline RL methods, sequence modeling-based methods, planning-based generative bidding methods, and LLM-based multi-modal methods. Traditional offline RL methods learn bidding policies by optimizing long-term rewards through value estimation and policy optimization~\cite{kumar2020conservative,kostrikovoffline,mou2022sustainable,wen2022cooperative}. For example, SORL~\cite{mou2022sustainable} combines offline pretraining with online RL to improve bidding stability, while MAAB~\cite{wen2022cooperative} models auto-bidding as a cooperative-competitive multi-agent optimization problem. Sequence modeling-based methods reformulate auto-bidding as a sequential prediction task and directly generate bidding actions from historical trajectories~\cite{chen2021decision, gao2025generative,lei2026generative,zhang2026generative}. For instance, GAS~\cite{li2025gas} enhances generative bidding through post-training search, while DRIVE~\cite{cui2026drive} leverages distributional modeling and retrieval augmentation to improve decision quality. Planning-based generative bidding methods further incorporate future planning mechanisms into generative policies to guide current bidding decisions~\cite{gao2025segb, guo2024generative, janner2022planning,li2025generative}. For example, DiffBid~\cite{guo2024generative} models bidding behaviors through conditional diffusion processes, while SEGB~\cite{gao2025segb} employs self-evolved autoregressive diffusion for long-horizon bidding planning. Recently, LLM-based multi-modal methods have attracted increasing attention due to their strong reasoning and language understanding capabilities. These methods leverage text information to model advertiser objectives and constraints while benefiting from the long-context reasoning ability of LLMs~\cite{lv2026decisionllm,li2026lbm,zhu2026role}. For example, LBM~\cite{li2026lbm} adopts a hierarchical reasoning-and-acting framework for bidding decision generation, while SemBid~\cite{zhu2026role} investigates the role of language representations in improving auto-bidding performance.\\
\textbf{LLM for sequential decision-making.}~~Recently, LLMs have been widely applied to sequential decision-making tasks such as web navigation~\cite{zhou2024webarena}, robotics~\cite{kimopenvla,intelligence2025pi_,pertsch2025fast}, and autonomous driving~\cite{li2025recogdrive, cui2023drivellm} due to their strong reasoning and semantic modeling capabilities. However, since LLM representations are inherently derived from discrete text modalities, they struggle to effectively model the continuous, high-dimensional, and dynamic state spaces and behavioral trajectories in sequential decision-making environments~\cite{dziri2023faith}. To enhance the capabilities of LLMs in sequential decision-making tasks, recent studies have adopted supervised fine-tuning methods that train models on manually annotated high-quality data, thereby improving their adaptability to complex environments. For example, APIGen-MT~\cite{prabhakar2026apigen} improves the adaptability of LLMs in multi-turn interaction and sequential decision-making tasks by constructing a simulated human-agent interaction framework to automatically generate high-quality multi-turn trajectory data for supervised fine-tuning. In addition, another line of research enhances the sequential decision-making capabilities of LLMs through RL, where PPO~\cite{schulman2017proximal} and its variants are widely adopted for policy optimization and long-term reward modeling~\cite{shao2024deepseekmath,feng2026group}. Some studies further incorporate modular designs, such as exploration~\cite{hao2026llm} and data augmentation~\cite{pang2024kalm,wan2025think}, to improve decision performance.
\subsection{Dataset Details}
\label{dataset}
\begin{table}[h]
\centering
\caption{Parameters of AuctionNet-dense and AuctionNet-Sparse.}
\label{tab:dataset}
\resizebox{\columnwidth}{!}{
\begin{tabular}{lcc}
\toprule
\textbf{Parameters} & \textbf{AuctionNet-dense} & \textbf{AuctionNet-sparse} \\
\midrule
Trajectories & 479,376 & 479,376 \\
Delivery periods & 9,987 & 9,987 \\
timestep in a trajectory & 48 & 48 \\
RTG dimension & 1 & 1 \\
State dimension & 16 & 16 \\
Action dimension & 1 & 1 \\
Action range & [0, 493] & [0, 589] \\
Impression's value range & [0, 1] & [0, 1] \\
CPA range & [6, 12] & [60, 130] \\
Total conversion range & [0, 1512] & [0, 57] \\
\bottomrule
\end{tabular}
}
\end{table}
In our experiments, we adopt AuctionNet~\cite{su2024a}, a large-scale public advertising bidding benchmark proposed and constructed by Alibaba. Specifically, the dataset provides two variants with different conversion sparsity levels, referred to in this paper as AuctionNet-dense and AuctionNet-sparse. Each variant contains over 500000 impression opportunities from 48 advertisers across 9987 delivery periods. Detailed dataset statistics are shown in Table \ref{tab:dataset}. In addition, the state at each timestep in a trajectory is constructed by aggregating historical statistics and current budget-related information. Specifically, each dimension represents the following information:
\begin{description}[leftmargin=0pt,labelsep=0.5em,font=\normalfont\bfseries]
    \item[time\_left:] The number of remaining decision intervals in the current advertising delivery period.
    
    \item[budget\_left:] The remaining budget available for allocation in the current delivery period.
    
    \item[historical\_bid\_mean:] The average bid submitted by the advertiser over all previous timesteps.
    
    \item[last\_three\_bid\_mean:] The moving average bid over the last three timesteps.
    
    \item[historical\_LeastWinningCost\_mean:] The average market clearing price over all previous timesteps, \ie the minimum cost required to win an auction.
    
    \item[last\_three\_LeastWinningCost\_mean:] The average market clearing price over the last three timesteps.
    
    \item[historical\_pValues\_mean:] The average predicted conversion probability over all previous timesteps.
    
    \item[last\_three\_pValues\_mean:] The average predicted conversion probability over the last three timesteps.
    
    \item[current\_pValues\_mean:] The average predicted conversion probability of impression opportunities arriving at the current timestep.
    
    \item[historical\_conversion\_mean:] The average number of conversions achieved over all previous timesteps.
    
    \item[last\_three\_conversion\_mean:] The average number of conversions over the last three timesteps.
    
    \item[historical\_xi\_mean:] The historical average winning rate, defined as the ratio between the number of won auctions and the total number of participated auctions.
    
    \item[last\_three\_xi\_mean:] The average winning rate over the last three timesteps.
    
    \item[current\_pv\_num:] The total number of available impression opportunities at the current timestep.
    
    \item[last\_three\_pv\_num\_total:] The cumulative number of impression opportunities over the last three timesteps.
    
    \item[historical\_pv\_num\_total:] The cumulative number of impression opportunities over all previous timesteps.
\end{description}

\begin{table}[h]
\centering
\caption{Detailed hyperparameter settings for SAGE.}
\label{tab:hyper}
\resizebox{0.7\columnwidth}{!}{
\begin{tabular}{lc}
\toprule
\textbf{Hyperparameter} & \textbf{Value} \\
\midrule
LLM backbone& Qwen2.5-Instruct-0.5B\\
Batch size & 128  \\
Learning rate & 1e-5  \\
Optimizer & AdamW  \\
Weight decay& 1e-4 \\
Grad clip & 0.25\\
Length of trajectory & 10  \\
Episode length & 48 \\
Total training steps & 400000 \\
Number of cross-attention head&8 \\
Number of experts & 4 \\
LoRA rank& 16  \\
Scaling factor&16 \\
\bottomrule
\end{tabular}
}
\end{table}

\subsection{Implementation Details}
\label{implementation}
Table \ref{tab:hyper} presents the detailed hyperparameter configurations of the proposed SAGE method. Notably, the hyperparameter settings remain largely consistent across different benchmark variants, which further demonstrates the robustness of our method.
\subsection{MoE Load Balancing Analysis}
\label{moe}
In our constraint-gated LoRA module, we do not explicitly introduce an auxiliary load-balancing loss. This design is motivated by the fact that each expert is intended to capture a distinct bidding strategy, and the strategy patterns are relatively separable under different constraint conditions. As a result, different tokens can be naturally routed to their corresponding experts based on the learned gating mechanism.

Nevertheless, to verify whether the routing process remains balanced in practice, we further report the expert utilization statistics during inference when the number of experts is set to 4. As shown in Table~\ref{tab:moe_balance}, the routing counts of the four experts are 445,752, 311,151, 435,567, and 324,570, respectively, corresponding to relatively balanced selection ratios of 29.38\%, 20.51\%, 28.71\%, and 21.39\%. These results indicate that the gating network does not collapse to a single expert, suggesting that the proposed constraint-gated LoRA module can effectively allocate different constraints to different experts without requiring an additional load balancing objective.
\begin{table}[h]
\centering
\caption{Expert utilization statistics during inference when the number of experts is set to 4.}
\label{tab:moe_balance}
\resizebox{0.9\columnwidth}{!}{
\begin{tabular}{lcc}
\toprule
\textbf{Expert} & \textbf{Selection Count} & \textbf{Selection Ratio} \\
\midrule
Expert 1 & 445,752 & 29.38\% \\
Expert 2 & 311,151 & 20.51\% \\
Expert 3 & 435,567 & 28.71\% \\
Expert 4 & 324,570 & 21.39\% \\
\bottomrule
\end{tabular}
}
\end{table}
\subsection{Text Prompt}
The task description prompt template and the condition constraint prompt template used in our experiments are provided as follows:

\begin{promptbox}{Task Description Prompt}

\small

You are an expert digital advertising strategist, embodied as an advanced auto-bidding algorithm.

\vspace{0.5em}

\textbf{Input Format}

You receive a fixed-length trajectory segment containing $K=10$ timesteps.

Each timestep $t$ is represented by three tokens in the following order:
\[
(RTG_t, S_t, A_t)
\]

where:
\begin{itemize}
    \item $RTG_t$: return-to-go at timestep $t$
    \item $S_t$: observed state at timestep $t$
    \item $A_t$: action taken at timestep $t$
\end{itemize}

\vspace{0.5em}

\textbf{Decision Objective}

At the current timestep $t$, predict the bidding multiplier $\alpha$ for that timestep.

\end{promptbox}


\begin{promptbox}{Condition Constraint Prompt}

\small

\textbf{Primary Goal}

Your primary goal is to maximize the total number of conversions for an ad campaign over a full day (the horizon), while strictly adhering to the budget and CPA constraints.

\vspace{0.5em}

\textbf{CPA Definition (Simple)}

CPA is defined as:
\[
\text{CPA} = \frac{\text{total\_spend}}{\max(\text{total\_conversions}, 1)}
\]

For decision-making at the current timestep, approximate the realized CPA using ONLY information observed so far in the trajectory (do NOT use future information).

\vspace{0.5em}

\textbf{CPA Reliability Principle}

Early in the campaign, when only a very small amount of budget has been spent or when conversions are very sparse, the realized CPA estimate is highly unreliable and should not dominate the bidding decision.

In such early stages:
\begin{itemize}
    \item Do not overreact to temporarily high CPA values caused by very small spend or zero conversions.
    \item Maintain a reasonable bidding level to gather sufficient impressions and conversions so that the CPA estimate becomes statistically meaningful.
\end{itemize}

As more budget is spent and more conversions are observed, gradually place more weight on the CPA constraint and adjust bids accordingly.

\vspace{0.5em}

\textbf{Guiding Principles}

You must follow these two core principles to achieve your goal:

\textbf{1. Dynamic CPA Management}
\begin{itemize}
\item If the realized CPA is reliably higher than the target after sufficient spending, decrease the bidding multiplier. However, when spending is still very small or conversions are sparse, avoid aggressively lowering bids based on unstable CPA estimates.
\end{itemize}

\textbf{2. Smooth Budget Pacing}
\begin{itemize}
\item Avoid spending the budget too quickly early in the day.
\item Aim for a smooth and consistent spending pattern across the entire horizon.
\item Avoid spending the budget too quickly early in the day, but also avoid bidding too conservatively when almost no budget has been spent.
\item Maintain sufficient participation in auctions to collect informative feedback signals.
\end{itemize}
\vspace{0.5em}

\textbf{Given Parameters}

\begin{itemize}
    \item Campaign Budget: \{budget\}
    \item Target CPA: \{cpa\_target\}
    \item Advertiser Category: \{category\}, category information may affect traffic quality, conversion efficiency, and bidding aggressiveness.
    \item Horizon: \{horizon\} timesteps (a day is split into 48 steps)
\end{itemize}

\end{promptbox}

\end{document}